\documentclass{mem}
\usepackage{natbib}\usepackage{txfonts}\usepackage{balance}
\usepackage{graphicx,epsfig}
\usepackage[a4paper]{hyperref}
\idline{0}{0}
\begin{document}
\def\teff{$T\rm_{eff }$}
\def\kms{$\mathrm {km s}^{-1}$}

\authorrunning{Mauro Dadina}
\titlerunning{The nearby Seyfert: from $BeppoSAX$ to $Simbol$-$X$}

\title{The Seyfert galaxies in the local Universe: from $BeppoSAX$ to $Simbol$-$X$}

\author{Mauro Dadina
\inst{1,}
\inst{2}}

\institute{$^{1}$INAF/IASF-Bo, via Gobetti 101, 40129 Bologna, Italy\\
$^2$Dipartimento di Astronomia dell'Universit\`a degli Studi di Bologna, 
via Ranzani 1, 40127 Bologna, Italy}


\date{Received date/ Accepted date}

\abstract{The operational conditions found by $BeppoSAX$ in observing nearby 
(z$\leq$0.1) Seyferts were reproduced for $Simbol$-$X$ in order 
to simulate a realistic final database of the mission. The results indicate 
that, even in the worst conditions, the $Simbol$-$X$ archive of pointings will 
allow to fully characterize the high-energy spectrum of nearby Seyferts and, 
most importantly, to obtain solid results on R and Ec 
(fundamental to model the cosmic X-Ray background, CXB). The measurement of 
the inclination angle of the accretion disk will be possible for $\sim$15 objects allowing to directly test 
the unified models for AGN. Finally, the time-dependent characteristics of the 
reflected component will be studied in at least $\sim$25 objects.



}

\maketitle

\vspace{-1.0cm}

\section{Introduction}

In X-ray astronomy each mission has been an important step forward in the 
comprehension of the physics of known sources and in the opening of new 
interesting fields of research.  This is expected to happen 
also for $Simbol$-$X$ (see, for instance the presentations by Fiore in 
these proceedings) thanks to its unprecedented sensitivity
above 10 keV ($\sim$2-3 orders of magnitude with respect the previous 
missions). Moreover $Simbol$-$X$ will also work as an observatory that will 
leave a rich but, as usual, inhomogeneous archive of data. 
As such,trying to estimate on how many objects 
deep spectral or timing studies will be performed, so as to fully exploit the 
astrophysical potentials of the new instruments, is not trivial. 
To make an attempt of tackling this issue for $Simbol$-$X$, it has been chosen 
to test the new mission reproducing the observational conditions 
(essentially the flux of the sources and the used exposure) really 
found by the only past mission comparable with $Simbol$-$X$: $BeppoSAX$. 
The astrophysical field inspected here is the one concerning the nearby 
(z$\leq$0.1) Seyfert galaxies, for which $BeppoSAX$ obtained important results 
on the X-rays emission mechanism, on the geometry of the reflector and on 
the geometry of the cold absorber (Perola et al. 2002, Risaliti et al. 1999). 
These results were obtained  investigating the properties of the high-energy 
cut-off (Ec), and of the reflection component (R) which study is possible only 
using broad-band observatories. These quantities are 
hardly measurable and they are known only for few bright objects. They were 
mainly measured from $BeppoSAX$ which operated above 10 keV with a smaller
($\sim$2-3 order of magnitude less) sensitivity with respect $Simbol$-$X$ but 
with a broader band ($\sim$0.1-200 keV, instead of $\sim$0.5-80 keV). The 
capability of inspecting Ec and R is here used as indicator of quality of the 
data assuming that, if R and Ec are measured, than it will be possible to 
deeply investigate, both via spectral and timing analysis, the 
physical conditions close the super-massive black-hole. The goal 
of this work is to estimate a realistic number of nearby Seyfert that
$Simbol$-$X$ will observe in this detail.

\scriptsize
\begin{table}
\caption{Grid of possible values for the interesting parameters of the simulated $Simbol$-$X$ spectra.}
\label{}
\vspace{-0.55cm}
\begin{center}
\begin{tabular}{cccccc}
\hline\hline
\\
N$_{H}^{a}$&$\Gamma$&R&Ec$^{b}$&F$_{2-10 keV}^{c}$&Exp.$^{d}$\\
&&&&&\\
\hline
&&&&&\\
0&&0.5&150&1&100\\
&&&&&\\
5&1.85$^{f}$&1&200&0.1&50\\
&&&&&\\
50&&1.5&250&0.01&10\\
&&&&&\\
\hline
\\
\end{tabular}
\end{center}
\vspace{-0.55cm}
$^{a}$ in units of 10$^{22}$ cm$^{-2}$; $^{b}$ in units of keV; $^{c}$ in units of 10$^{-11}$ erg s$^{-1}$ cm$^{-2}$; $^{d}$ in units of ks; $^{f}$ fixed 

\end{table}
\normalsize

\vspace{-0.35cm}

\section{Methods}

\vspace{-0.08cm}

The observational conditions (exposure time end fluxes of the sources) found
by $BeppoSAX$ were grossly reproduced in order to simulate the results
that would have been obtained if the nearby Seyferts were observed
with $Simbol$-$X$ instead of $BeppoSAX$. To do that, the parameter F defined 
as follows, has been used:

\vspace{0.2cm}

F=(F$_{2-10 keV}$$\times$$\sqrt{T_{exp}}$) \hspace{1cm}(1)

\vspace{0.2cm}

where F$_{2-10 keV}$ is the 2-10 keV flux in units of 10$^{-11}$ erg s$^{-1}$ 
cm$^{-2}$ and T$_{exp}$ is the exposure time in units of ks. Thus, the 
F parameter define how ``deep'' an observation was with $BeppoSAX$ 
and may be used to quantify the expected improvements achievable with the 
new instruments. The distribution of F obtained for  
$BeppoSAX$ is presented in figure 1 (note that Log(F)=1 
for 100 ks long observations of sources having F$_{2-10 keV}$=10$^{-11}$ erg 
s$^{-1}$ cm$^{-2}$).
The $BeppoSAX$ catalog used here is the one presented in 
Dadina (2007), which contains 163 observations of 105 objects (of which 43 are 
type I and 62 are type II Seyferts). 

To produce the $Simbol$-$X$ fake spectra it was assumed that the X-ray 
emission of the sources is described by an absorbed 
(by cold matter) power-law with an Ec, plus a cold reflection component and a 
narrow emission line. 
Each combination of the values presented in Table 1 has been 
used to simulate the fake $Simbol$-$X$ spectra. When R and Ec were available 
from $BeppoSAX$, the closest value between the 3 reported in table 1 were used 
for fake data, otherwise it was set R=1 and Ec=200 keV. The resulting grid of 
spectra have been fit with what presented in Dadina (2007) in order to
reproduce for $Simbol$-$X$ the distribution of F presented in figure 1. 
Finally 90\% cofidence levels have been obtained for each interesting 
parameters in the simulated spectra.

\begin{figure}
\includegraphics[width=6.8cm,height=2.5cm,angle=0]{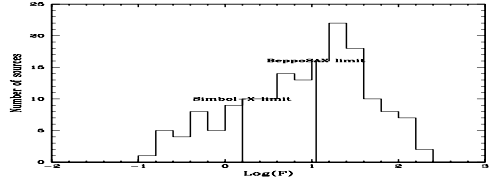}
\caption{Histogram of the real distributions of the F parameters (see text for details) obtained for the $BeppoSAX$ list of observations of the nearby Seyferts. The vertical line at F$\sim$1 is the limit reached for $BeppoSAX$ in 
inspecting R and Ec, while the one at F$\sim$0.2 is the expected limit for $Simbol$-$X$.}

\vspace{-0.3cm}
\end{figure}

The configuration assumed for $Simbol$-$X$ is 
characterized by the following specifications:

\vspace{-0.2cm}

\begin{itemize}

\item Internal particle background for the low energy detector MPD = 2$\times$10$^{-4}$cts/s/keV

\item Internal particle background for the high energy detector CZT = 3$\times$10$^{-4}$cts/s/keV

\item MPD dead-time = 17\%

\end{itemize}

\vspace{-0.2cm}
 
 as specified in the tutorial for the $Simbol$-$X$ simulations (Sauvageot, 
2007).

\vspace{-0.3cm}

\section{$Simbol$-$X$ vs. $BeppoSAX$}

The first remarkable results obtainable with $Simbol$-$X$ is that all the
pointed nearby Seyfert galaxies observed with $BeppoSAX$ are expected to be 
detected above 10 keV with respect $\sim$80\% detected by $BeppoSAX$.
At the same time the range of F parameters available for the inspection of
the properties of R and Ec increases by $\sim$1 order of magnitude 
(see figure 1).
The high sensitivity expected in the range 10-70 keV will allow to study Ec 
even well above 70 keV (see figure 2), i.e. at energies exceeding 
the upper end of the nominal working range of the CZT. 
Nonetheless, the possibility to detect Ec at energies above 200 keV is 
limited to brighter sources (see lower panel of figure 2).

\begin{figure}
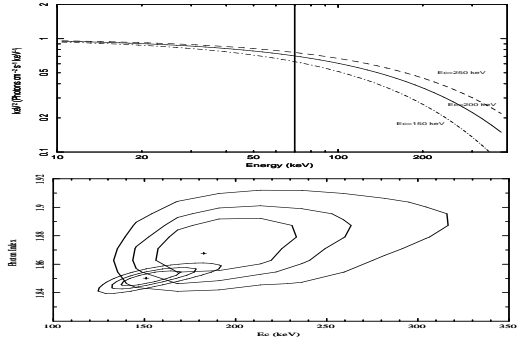

\includegraphics[width=2.2cm,height=6.6cm,angle=270]{dadina_f2.ps}
\includegraphics[width=2.2cm,height=6.8cm,angle=270]{dadina_f3.ps}
\caption{Upper panel: 10-200 keV spectra for power-law with Ec=150 keV (lower curve), Ec=200 keV (middle curve), and Ec=250 keV (upper curve). It is worth noting that at 70 keV the three line differ for $\sim$10\% each other and that the difference decreses as Ec= increases. Lower panel: $\Gamma$ vs. Ec confidence contours for a F=1, R=1 observation for Ec=150 keV (smaller contours) and Ec=200 keV (bigger contours).}
\vspace{-0.3cm}
\end{figure}

The number of Seyfert for which we can have information 
(both detections and reliable lower limits) about Ec and R will double 
(going from 41 to 82) assuming the $BeppoSAX$ archive as the reference one. 
Most interestingly, a significant  fraction of the ``new'' sources (26 out of 
41) for which information about Ec and R will be for the first time achievable 
are type  II objects. These were almost excluded in the $BeppoSAX$  era 
since the spectral complexity due to the cold absorption. The coupling 
offered by $Simbol$-$X$ of a good sensitivity below 10 keV (comparable to the one of $XMM$-$Newton$  or $Suzaku$) and the great improvement 
($\sim$two orders of magnitude if compared  with $BeppoSAX$ and $Suzaku$) 
between 10-70 keV should at least partially disentangle this degeneracy.  

$Simbol$-$X$ is expected to give a major contribute in the measurements
of the reflection characteristics in nearby Seyfert galaxies. In fact, the
reflection hump peaks at 30-40 keV (Lightman \& White 1987), i.e. just in the 
middle of the working range of the CZT 
detector. Moreover, $Simbol$-$X$ will permit to investigate
the time-dependent behavior of the reflection. In figure 3,
the confidence contours are plotted for three different observations of
a Seyfert 1 (upper panel) with F=1 of that changes the value of R from 
R=0.5 to R=1.5. In the lower panel of the same figure, the contour plot for a
type II source (F=1, R=1, Ec=200 keV) is presented and shows how the presence 
of the cold absorption prevent this kind of studies.

\begin{figure}
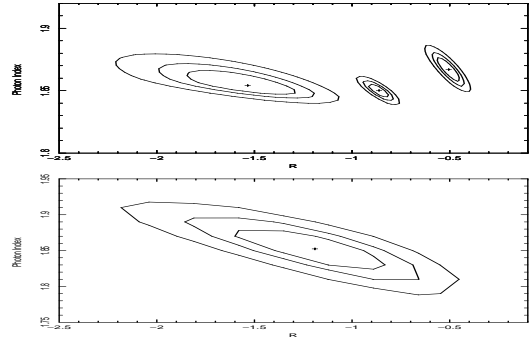

\includegraphics[width=2.2cm,height=6.8cm,angle=270]{dadina_f4.ps}
\includegraphics[width=2.2cm,height=6.8cm,angle=270]{dadina_f5.ps}
\caption{Upper panel:confidence contours for R going from 0.5 to 1.5 
for a simulated type I objects (F=1, Ec=200 keV). Lower panel: confidence 
contours for a simulated Seyfert II with R=1 (F=1, Ec=200 keV).}
\vspace{-0.4cm}
\end{figure}

For very bright (F$\geq$10) sources, $Simbol$-$X$ is expected to 
allow the study of the inclination angle of the accretion disk
(figure 4). This kind of studies will
be permitted only in type I sources since the spectral complexity 
introduced by the absorber at low energy will hamper to constraint this 
parameter.

\vspace{-0.2cm}

\section{Summary and conclusions}

This work is a first attempt to determine a realistic number
of nearby (z$\leq$0.1) Seyfert galaxies for which $Simbol$-$X$ will 
carry deep spectral and timing analysis. The basic assumption here are: 
1) that these studies will be possible when the measurements of R and Ec are 
possible; 2) that Simbol-X will observe sources  with a distribution of F 
similar to what done by $BeppoSAX$. The last assumption is quite strong and
it is probable that the characteristics of Simbol-X will be used to inspect 
sources that were barely detectable with $BepoSAX$, as the Compton-thick 
sourcces. Thus, what obtained here may be considered as lower limits.   
The results can be summarized as  follows: i) the fraction of nearby Seyfert 
galaxies detected above 10 keV should
be $\sim$100\%; ii) for $\sim$80 Seyfert galaxies in the local universe it 
will be possible to measure Ec and R; iii) $\sim$50\% of these sources are 
expected to be type II sources;  iv) for $\sim$25 objects it will be possible 
to perform meaningful time dependent studies of R; v) for $\sim$15 type I 
sources is will be possible to study the inclination angle of the accretion 
disk inspecting the properties of the reflected components.

\begin{figure}
\includegraphics[width=2.2cm,height=6.4cm,angle=270]{dadina_f6.ps}
\includegraphics[width=6.8cm,height=2.2cm,angle=0]{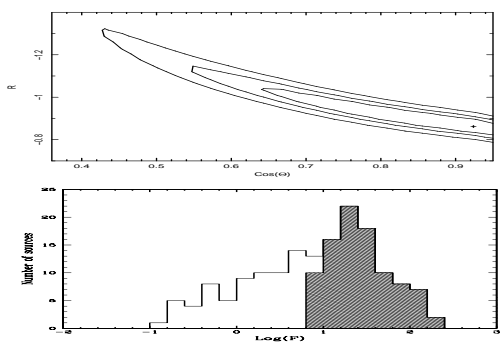}
\caption{Upper panel: Confidence contours for the inclination angle ($\Theta$)
for a type I source with F=1, R=1 
and Ec=200. Lower panel: histogram of the F parameter. The shaded region 
correspond to the observations for which information on the inclination angle 
is available studying the reflected component.}
\vspace{-0.3cm}
\end{figure}

These results, thus, indicate that $Simbol$-$X$ is expected to leave an 
archive of pointed observations that will allow to fully characterize 
the average spectrum of nearby Seyfert galaxies on solid statistical basis 
(for instance, the distribution of N$_H$ for the Seyfert galaxies in the local 
Universe was obtained with a sample of 74 objects, Risaliti et al. 1999). Most 
importantly, it is expected to obtain solid description of the distributions 
of R and Ec, fundamental to synthesize the CXB. Moreover it will be possible  
to test the UM for the brightest Seyferts checking if the 
inclination angle of the system is in accordance with the optical 
classification of the sources. Finally, the astrophysics of
the accretion will be studied using timing techniques at the energies where 
the bulk of the reflection peaks.

\vspace{-0.45cm}

\subsection{Caveats}

\vspace{-0.2cm}

Some basic assumption and simplifications have been 
made in this work and these must be kept in mind when evaluating 
the the opportunities offered by $Simbol$-$X$ in the inspected astrophysical 
field. The sample of objects/observations used here have been 
accumulated in six years while $Simbol$-$X$ is expected to fly 
for 2-5 years (Ferrando et al., 2006). Nonetheless, the higher observational 
efficiency (up to $\sim$90\%, Ferrando et al., 2006) should permit 
$Simbol$-$X$ to observe the nearby Seyferts for $\sim$10 Ms as 
BeppoSAX did. The baseline model assumed here does not include warm absorber 
or ionized reflection, nor include soft excess in type I sources. 
All these components introduce additional spectral complexities not accounted 
for. On the contrary, the capabilities of $Simbol$-$X$ in measuring R and Ec 
were here tested only on pure statistical basis: i.e. no physical assumption 
were made during the fitting and the significance of the measurements of R and 
Ec were calculated leaving all the other parameters free to vary.

\vspace{-0.2cm}

\small
\begin{acknowledgements}
Financial support from ASI under contract ASI/INAF I/023/05/0 is aknowldeged
\end{acknowledgements}

\end{document}